\documentclass[twocolumn]{aastex631}

\usepackage{chngpage}
\usepackage{booktabs}
\usepackage{amsmath}
\usepackage{xcolor}
\usepackage{bm}

\newcommand\RGMX{\bgroup\markoverwith{\textcolor{cyan}{\rule[0.5ex]{4pt}{1pt}}}\ULon}

\shorttitle{GR Effects on Circumbinary Disks}
\shortauthors{Childs et al.}

\begin{document}

\title{Relativistic Effects on Circumbinary Disk Evolution: Breaking the Polar Alignment around Eccentric Black Hole Binary Systems}

\author[0000-0002-9343-8612]{Anna C. Childs}
\affiliation{Center for Interdiskiplinary Exploration and Research in Astrophysics (CIERA) and Department of Physics and Astronomy Northwestern University,1800 Sherman Ave, Evanston, IL 60201 USA}
\author[0000-0003-2401-7168]{Rebecca G. Martin}
\affiliation{Nevada Center for Astrophysics, University of Nevada, Las Vegas, NV 89154, USA}
\affiliation{Department of Physics and Astronomy, University of Nevada, Las Vegas, 4505 South Maryland Parkway,
Las Vegas, NV 89154, USA}

\author[0000-0002-2137-4146]{C. J. Nixon}
\affiliation{School of Physics and Astronomy, Sir William Henry Bragg Building, Woodhouse Ln., University of Leeds, Leeds LS2 9JT, UK}

\author[0000-0002-3881-9332]{Aaron M. Geller}
\affiliation{Center for Interdiskiplinary Exploration and Research in Astrophysics (CIERA) and Department of Physics and Astronomy Northwestern University,1800 Sherman Ave, Evanston, IL 60201 USA}

\author[0000-0002-4636-7348]{Stephen H. Lubow}
\affiliation{Space Telescope Science Institute, 3700 San Martin Drive, Baltimore, MD 21218, USA}

\author[0000-0003-3616-6822]{Zhaohuan Zhu}
\affiliation{Nevada Center for Astrophysics, University of Nevada, Las Vegas, NV 89154, USA}
\affiliation{Department of Physics and Astronomy, University of Nevada, Las Vegas, 4505 South Maryland Parkway,
Las Vegas, NV 89154, USA}

\author[0000-0003-2270-1310]{Stephen Lepp}
\affiliation{Nevada Center for Astrophysics, University of Nevada, Las Vegas, NV 89154, USA}
\affiliation{Department of Physics and Astronomy, University of Nevada, Las Vegas, 4505 South Maryland Parkway,
Las Vegas, NV 89154, USA}



\begin{abstract}
We study the effects of general relativity (GR) on the evolution and alignment of circumbinary disks around binaries on all scales. We implement relativistic apsidal precession of the binary into the hydrodynamics code {\sc phantom}.  We find that the effects of GR can suppress the stable polar alignment of a circumbinary disk, depending on how the relativistic binary apsidal precession timescale compares to the disk nodal precession timescale. Studies of circumbinary disk evolution typically ignore the effects of GR which is an appropriate simplification for low mass or widely separated binary systems. In this case, polar alignment occurs providing that the disks initial misalignment is sufficiently large.  However, systems with a very short relativistic precession timescale cannot polar align and instead move toward coplanar alignment. In the intermediate regime where the timescales are similar, the outcome depends upon the properties of the disk. Polar alignment is more likely in the wavelike disk regime (where the disk viscosity parameter is less than the aspect ratio, $\alpha<H/r$) since the disk is in good radial communication. In the viscous disk regime disk breaking is more likely. Multiple rings can destructively interact with one another resulting in short disk lifetimes, and the disk moving towards coplanar alignment.
Around main-sequence star or stellar mass black hole binaries, polar alignment may be suppressed far from the binary but in general the inner parts of the disk can align to polar. Polar alignment may be completely suppressed for disks around supermassive black holes for close binary separations.
\end{abstract}


\section{Introduction}

Misaligned circumbinary disks are observed or inferred to exist around binaries on a wide range of scales from main-sequence (MS) stars up to supermassive black hole (SMBH) binaries.
Around MS stars, many circumbinary gas and debris disks that are misaligned to the binary orbital plane have been observed \citep[e.g.][]{Chiang2004,Capelo2012,Brinch2016}.  Disk misalignment may happen as a natural consequence of a chaotic formation process \citep{Bate2010,Offner2010,Bate2018}, or later in time from stellar encounters \citep{Clarke1993}.
Furthermore, two polar aligned gas disks \citep{Kennedy2019,Kenworthy2022} and one debris disk \citep{Kennedy2012} have been observed that are inclined by $90^{\circ}$ to the binary orbital plane.  Stellar-mass black hole binaries may host misaligned circumbinary disks as a result of chaotic accretion. In particular,  binaries in active galactic nucleus (AGN) disks may be likely to host inclined circumbinary disks \citep{Li2022,Li2022b}.

There is evidence that SMBH binaries may lie at the center of many AGN \citep{DeRosa2019}.  The gravitational wave background detected recently by NANOGrav is thought to originate due to the emission from SMBH binaries \citep[e.g.][]{Agazie2023}.  Furthermore, there is tentative evidence of short-period SMBH eccentric binaries (in the process of merging) at the centers of AGN \citep{Jiang2022}.  SMBH binaries that form from mergers are thought to typically be eccentric \citep{Gualandris2022}.  Interactions with prograde circumbinary disks can grow the eccentricity if the eccentricity is initially small \citep{Roedig2011} and similarly, interactions with retrograde circumbinary disks can increase the binary eccentricity \citep{Nixon2011,Nixon2015}.  Thus, we expect SMBH binaries to form eccentric and remain eccentric until they reach the gravitational wave regime \citep[at which point their eccentricity diminishes faster than their semi-major axis and they merge with an orbit that is close to circular,][]{Begelman1980}.

Observations of jet directions suggest that large disk misalignments to the host galaxy axis are possible \citep[e.g.][]{Kinney2000}.  Such misalignment indicates that the spin direction of the SMBH is not aligned with the galaxy, and thus the accretion history into the central regions of the galaxy is not likely to be in any preferential plane \citep{King2006, King2007}. This “chaotic accretion” scenario has also been applied to the last parsec problem to help merge SMBHs \citep[e.g.][]{Nixon2011, Nixon2011a, Nixon:2013}.  Evidence of warped accretion disks around SMBH in active galactic nuclei (AGN) has been available from e.g. water maser emission \citep{Herrnstein1996}. More recently it has been possible to constrain the inclination of AGN disks through modelling of double-peaked emission line profiles observed from galaxy nuclei \citep{Dias2023}. Such modelling indicates that detecting highly inclined accretion disks may be possible. 

Theoretical studies have shown that misaligned disks can evolve either towards a coplanar alignment or a polar configuration around eccentric binaries \citep{Aly2015,Martin2017,Martin2018,Lubow2018,Zanazzi2018,Cuello2019,Ceppi2023,Rabago2023}.
This is because particle orbits around an eccentric binary undergo nodal precession, either about the binary angular momentum vector (circulating orbits) or the stationary inclination (librating orbits), depending upon their initial inclination \citep{Verrier2009,Farago2010,Doolin2011,Naoz2017,Chen2019}. 
Consider the case of circumbinary test particle that is sufficiently far from the binary that the binary potential
can be accurately approximated by an orbit averaged (secular) potential.  In the absence of general relativity (GR), the particle orbit does not precess (is stationary) at an orientation with
its inclination perpendicular to the binary orbital plane and its angular momementum parallel
to the binary eccentricity vector. The conditions for a stationary orbit are independent of orbital radius.

However, there exists a radial limit for stationary polar circumbinary orbits that arises as a result of the relativistic apsidal precession of the binary \citep{Lepp2022}.  Particles close to the binary precess with the binary and are able to maintain their polar librating orbits \citep{Childs2021}. Farther away from the binary, particle orbits decouple from the motion of the binary and the stationary inclination and the minimum inclination required for libration increases with particle separation.  


Most theoretical studies of circumbinary disk evolution ignore the effects of GR.  For most low mass and wide binaries, relativistic effects are negligible close to the binary. However, GR effects may be important in some parameter space, even for low mass stellar binaries \citep{Zanardi2017,Lepp2022}. 
Here, we show that significant relativistic apsidal precession of the binary can inhibit the polar alignment of a circumbinary disk.  We investigate the effects of GR on circumbinary disk evolution by running a suite of smoothed particle hydrodynamic (SPH) simulations using code that has been modified to include the effects of GR.  In Section~\ref{sec:discussion} we discuss the theoretical framework which motivates this study and predicts when GR effects become important in driving the disk evolution around various binary systems on a range of scales. In Section \ref{sec:Simulations} we describe our modifications to the SPH code and our simulation setups.  In Section \ref{sec:results} we present our simulation results.  Lastly, we summarize our findings in Section \ref{sec:conclusions}.

\section{When is GR important?}\label{sec:discussion}

The timescale for the prograde binary apsidal precession from GR is
\begin{equation}
t_{\rm GR}=2 \pi \frac{a_{\rm b}^{5/2}c^2(1-e_{\rm b}^2)}{3 (GM_{\rm b})^{3/2}},
\end{equation}
where $M_{\rm b}$ is the total mass of the binary
\citep{Zanardi2018}.
As a result of this binary precession, polar circumbinary orbits do not exist beyond a critical radius given by
\begin{equation}
    \frac{r_{\rm c}}{ a_{\rm b}}= \left ( \frac{a_{\rm b}c^2M_1M_2}{8G(M_1+M_2)^3} (1-e_{\rm b}^2)(2+8e_{\rm b}^2) \right )^{2/7}
    \label{rc}
\end{equation}
\citep{Lepp2022},
for a binary with separation $a_{\rm b}$ and eccentricity $e_{\rm b}$,
where $G$ is the gravitational constant, $M_1$ and $M_2$ are the masses of the primary and secondary binary components respectively, and $c$ is the speed of light. 
  The critical radius given in equation~(\ref{rc}) describes the radius inside of which librating orbits exist. Polar alignment of a disk is therefore only possible inside of this radius. If $r_{\rm c}$ is much larger than the outer radius of the disk, then GR cannot affect the circumbinary disk evolution. The destabilizing effects of GR become important when 
  $r_{\rm c}$ is within the radial range of the disk. So for what binary parameters do GR effects need to be considered? 

\begin{figure}
	\includegraphics[width=1\columnwidth]{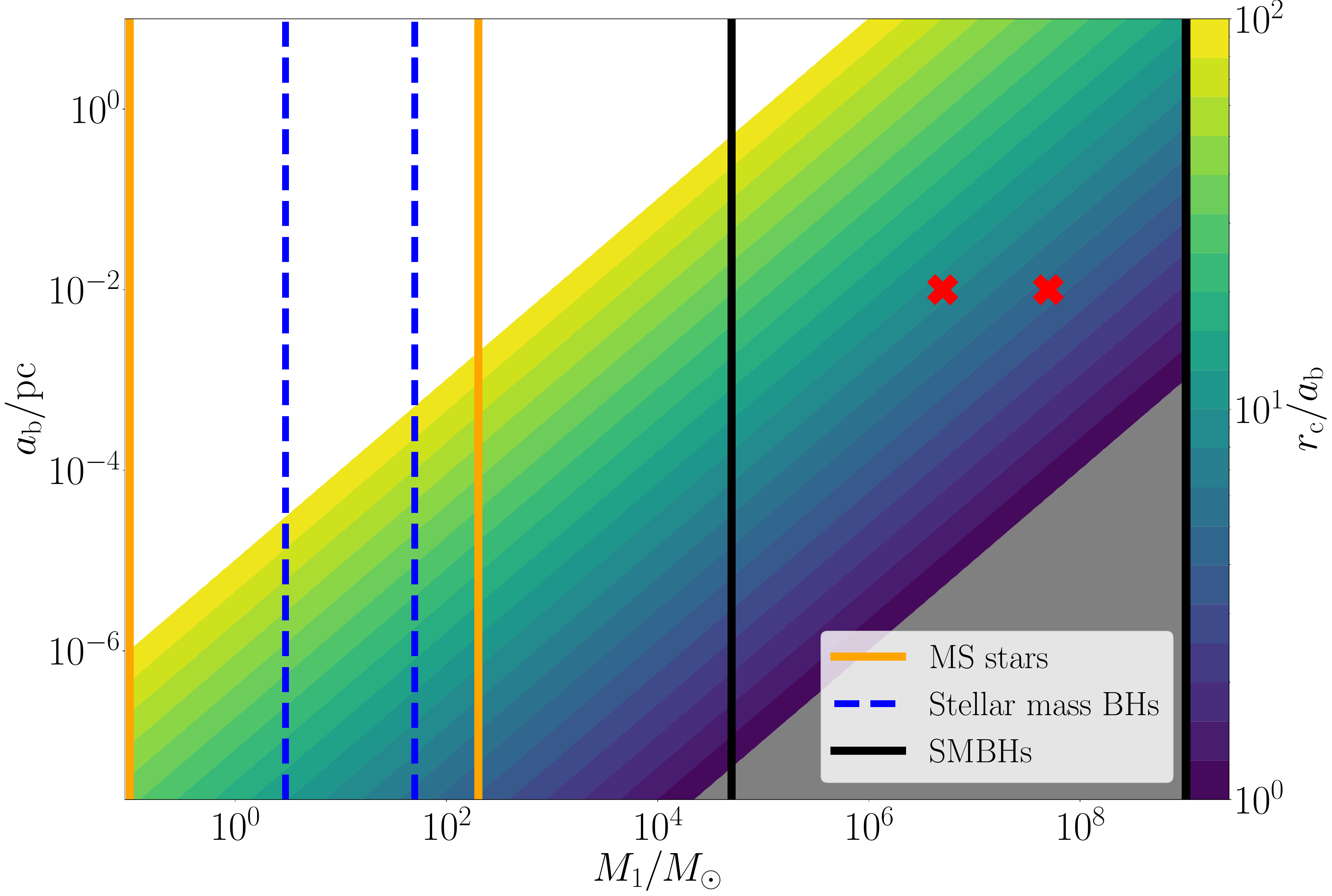}
    \caption{Contour map of $r_{\rm c}/a_{\rm b}$ values for various binary primary mass, $M_1$, and separation, $a_{\rm b}$, values.  The binary is equal mass with $e_{\rm b}=0.5$.  The gray region is where $r_{\rm c}/a_{\rm b}<1$ and the white region is where $r_{\rm c}/a_{\rm b}>100$.  The two red crosses marks the parameter space we sample in our simulations The vertical lines mark the mass regimes for various binary systems.  The binary-disk evolution scales with the value of $r_{\rm c}/a_{\rm b}$ such that the simulations we run are applicable to systems on a wide range of scales.}
    \label{fig:rc}
\end{figure}

Fig.~\ref{fig:rc} shows a contour map of $r_{\rm c}/a_{\rm b}$ values for various binary primary mass $M_1$ and separation values $a_{\rm b}$.  We set $e_{\rm b}=0.5$  and the binary mass ratio equal to 1 to calculate $r_{\rm c}$. \cite{Lepp2022} found that $r_{\rm c}$ is not sensitive to binary eccentricity unless that eccentricity approaches one.  The gray region marks where $r_{\rm c}/a_{\rm b}<1$ and the critical radius is smaller than the binary separation. For binaries in this region, there are no librating orbits around them and therefore polar disk alignment is not possible at any radius around the binary.  The white region marks where $r_{\rm c}/a_{\rm b}>100$. In this region the timescale for the GR apsidal precession is long and the effects of GR become negligible. The colored region shows binary parameters for which GR may affect the disk evolution, depending upon the properties of the disk.  The two red crosses marks the parameter space we sample in our simulations in Section~\ref{sec:results}.

From equation~(\ref{rc})  we see that $r_{\rm c}/a_{\rm b} \propto  (a_{\rm b}/M_{\rm b})^{2/7}$ for an equal mass binary. Systems with the same value of  $r_{\rm c}/a_{\rm b}$ have the same dynamics (i.e. the disk evolution is the same if the ratio $M_{\rm b}/a_{\rm b}$ is fixed).
While the simulations we consider in Section \ref{sec:results} are for masses relevant for SMBHs, the same behaviour is expected for lower mass objects that are closer together.     For example, we have run simulations with $M_{\rm b}=10^7\,\rm M_\odot$ and $a_{\rm b}=0.01\,\rm pc$ but the same behaviour (same $r_{\rm c}/a_{\rm b}$) is expected for say $M_{\rm b}=100\,\rm M_\odot$ and $a_{\rm b}=1\times 10^{-7}\,{\rm pc}=4.4\,\rm R_\odot$.  For a binary comprised of two Sun-like stars, this $r_{\rm c}/a_{\rm b}$ requires an $a_{\rm b}$ value much less than the radii of the stars and is thus not possible.

Fig. \ref{fig:rc} highlights the parameter space relevant for binaries on different scales. The vertical orange lines mark the approximate mass range of MS stars ($0.1-200 \, M_{\odot}$).  In this range we find $r_{\rm c}/a_{\rm b}$ values larger than what we are able to simulate. For example, with stars of mass $M_1=M_2=1\,\rm M_\odot$ with a semi-major axis of $a_{\rm b}=0.2\,\rm au$ and eccentricity $e_{\rm b}=0.5$, we find $r_{\rm c}=51.5\,a_{\rm b}$. This critical radius could be in the radial range of a protoplanetary disk if the disk is sufficiently radially extended. For these parameters, however, we find that $t_{\rm GR}=1.6\times 10^5{\,\rm yr}=2.5\times 10^6\,T_{\rm b}$, where $T_{\rm b}$ is the binary orbital period. While this is much shorter than the expected disk lifetime, it is  prohibitively long for running a disk simulation for this many binary orbits.
Within the MS star mass range we also show the range of stellar mass BHs ($3-50 \, M_{\odot}$) marked by vertical dashed blue lines.  Similarly to protoplanetary disks around MS stars, the timescales involved for the GR precession are too long for us to simulate.

The approximate SMBH mass range is marked by vertical black lines ($5 \times 10^4-1 \times 10^9 \, M_{\odot}$) and within this mass range we find much smaller $r_{\rm c}/a_{\rm b}$ values are possible. For such  systems, the timescale for relativistic precession is shorter in units of the binary orbital period, allowing us to get more timely simulation results.   





\section{Simulations}\label{sec:Simulations}

\begin{figure*}[!ht]
	\includegraphics[width=2\columnwidth]{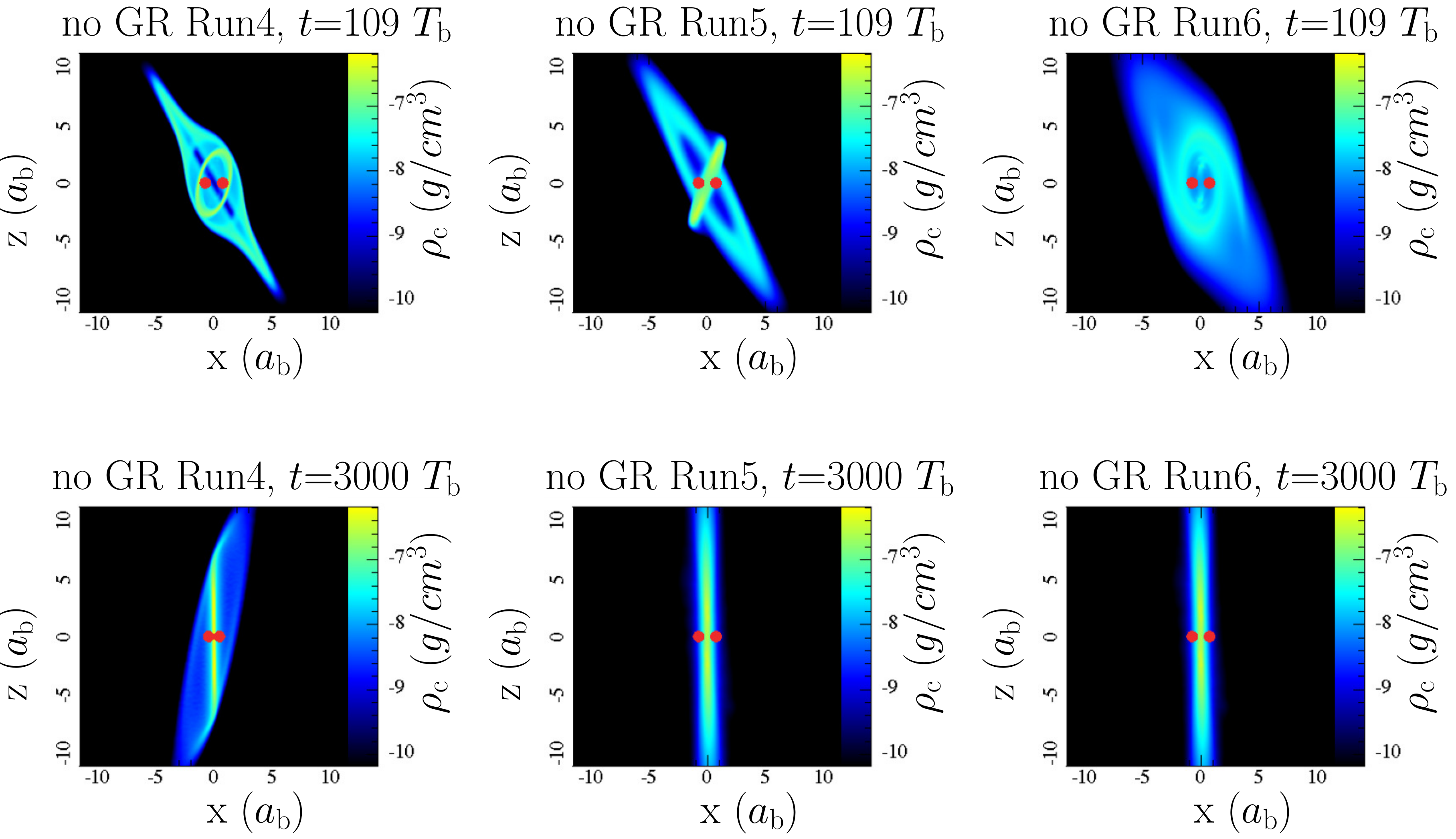}
    \caption{Column density images shown in the $x-z$ plane for the runs without GR at two different times. The binary orbits in the $x-y$ plane and is shown by two red dots.  The binary eccentricity vector is along the $x$-axis.}
    \label{fig:nogr_splash}
\end{figure*}

We use the SPH \citep[][]{Monaghan1992,Price2012a} code {\sc phantom} \citep{Lodato2010,Price2010,Price2018}  that we have modified to include the effects of GR. The binary components are treated as sink particles that accrete the mass and angular momentum of any particles that fall inside their sink radius \citep{Bateetal1995}.  To implement relativistic precession of the binary we follow \cite{Nelson2000} and modify the acceleration of each sink particle as a result of sink-sink interaction. 

The binary components 1 and 2 are located at position vectors $\bm{r}_1$ and $\bm{r}_2$, respectively, and have the relative coordinate vector $\bm{r}=\bm{r}_2-\bm{r}_1$.  The gravitational force per unit mass on sink 1 is
\begin{equation}
    -  {\bf F}_1 =  \frac{GM_{2}}{r^3}\left(  1 + \frac{6R_+}{r}  \right ) \textbf{r},
\end{equation} 
and similarly  the force on sink 2 is
\begin{equation}
    -  {\bf F}_2 =  - {\bf\nabla}  {\bf F}_1 \frac{M_1}{M_2},
\end{equation} 
where   $R_{+}=GM_{\rm b}/c^2$,  $M_{\rm b}=M_1+M_2$ is the total mass of the binary and $r=|\bm{r_2}-\bm{r_1}|$ is the instantaneous separation of the binary components.
While this force differs from the commonly used pseudo-Newtonian expression of \cite{Paczynsky1980} that gives the correct radius of the last stable circular orbit, this gives the correct orbital apsidal precession frequency. 

For the first set of parameters we choose $M_1=M_2=5\times 10^6\,\rm M_\odot$ and $a_{\rm b}=0.01\,\rm pc$, so that $r_{\rm c}=8.8 \,a_{\rm b}$ and $t_{\rm GR}=1.5\times 10^5{\,\rm yr}=5215\,P_{\rm orb}$. For the second set of parameters we increase the masses to $M_1=M_2=5\times 10^7\,\rm M_\odot$, which corresponds to $r_{\rm c}=4.6 \,a_{\rm b}$ and $t_{\rm GR}=522\,P_{\rm orb}$.  This more massive system has a similar
mass and orbital radius as candidate SDSS J0159+0105 \citep{Zheng2016}. In Section \ref{sec:discussion} we showed that the results are scale free with a fixed ratio of $a_{\rm b}/M_{\rm b}$.
We set the binary eccentricity to $e_{\rm b}=0.5$.  The changes in $a_{\rm b}$ and $e_{\rm b}$ due to GR over time $t_{\rm GR}$ are negligible.  Each compact object has an accretion radius of $0.25 \, a_{\rm b}$.  

\begin{table}\label{tab:runs}
\centering
\caption{Simulation setups.  Runs with an asterisk indicate simulations were run both with and without GR for this setup.  All binaries are equal mass and separated by $a_{\rm b}=0.01 \, \rm pc$.  The disk has an $\alpha$ viscosity value of 0.05 and is initially tilted by $60^{\circ}$ to the binary orbital plane.}
\begin{tabular}{cccc}
  \hline
Simulation name  & {$M_{\rm 1}/M_{\odot}$} & {$H/r$} & {$r_{\rm c}/a_{\rm b}$} \\
\hline
\hline
Run1 & $5 \times 10^6$ & {0.02} & 8.8 \\
Run2 & $5 \times 10^6$ &{0.05} & 8.8 \\
Run3 & $5 \times 10^6$ &{0.10} & 8.8 \\
Run4* & $5 \times 10^7$ & {0.02} & 4.6 \\
Run5* & $5 \times 10^7$ & {0.05} & 4.6 \\
Run6* & $5 \times 10^7$ &  {0.10} & 4.6 \\
\hline
\end{tabular}
\end{table}

We are interested in the evolution of a  disk  with a mass that is small compared to the binary mass so that the disk mass does not affect the binary evolution.  Thus, the circumbinary disk is initialized with a small mass of $1\times10^{-6} \, M_{\odot}$.  The disk mass is small enough that it does not affect the binary motion. The same evolution is expected independent of the disk mass provided that its angular momentum remains small compared with the binary angular momentum.  The disk is composed of 500,000 particles that are distributed in the radial range $r_{\rm in}=2 \, a_{\rm b}$ to $r_{\rm out}=10 \, a_{\rm b}$.  The surface density profile of the gas follows
\begin{equation}
    \Sigma(r) = \Sigma_0R^{-1.5}
\end{equation}
and the sound speed profile follows
\begin{equation}
    c_{\rm s} = c_{\rm s0}R^{-3/4},
\end{equation}
where $\Sigma_0$ and $c_{\rm s0}$ are scaling constants. These choices for the power laws allow the \citet{SS1973} $\alpha$ parameter and the shell averaged smoothing length divided by the disk scale height, $\left<h\right>/H$,  to be constant over the disk radius \citep{Lodato2007}.
The disk is initially tilted by $60^{\circ}$ relative to the binary orbital plane in a direction towards the binary eccentricity vector \citep[with longitude of ascending node relative to the binary eccentricity vector of $\phi=90^\circ$, see equation 3 of ][]{Chen2020}. In the absence of GR effects, such a disk is expected to undergo nodal libration and align to polar \citep{Martin2017,Lubow2018}.

We experiment with different disk aspect ratios ($H/r$).   The $H/r$ values are taken at the initial disk inner radius $r_{\rm in}=2 \, a_{\rm b}$.  The $\alpha$ viscosity is implemented by adapting the SPH artificial viscosity with $\alpha_{\rm AV} = 0.624$ such that $\alpha = 0.05$ \citep[e.g.][]{Lodato2010}, and to resolve shocks that form in the disk we add a quadratic viscosity with $\beta=2$.  The disk is initially resolved with a mean smoothing length over disk scale height of $\left<h\right>/H=0.80,0.43,\, \rm and \, 0.27$ for disks with $H/r=0.02,0.05, \, \rm and \, 0.1$ respectively.  While $H/r$ is likely lower in the inner regions of AGN disks \citep{Cantiello2021}, running lower aspect ratios would be very computationally expensive. However,  these choices of disk aspect ratio allow us to consider the viscous disk regime ($\alpha>H/r=0.02$), the intermediate regime ($\alpha=H/r$) and the wave-like disk regime ($\alpha<H/r=0.1$).  

Table~\ref{tab:runs} lists the different setups we experiment with.  In Runs 4, 5, and 6, the runs with asterisks in Table~\ref{tab:runs}, we perform SPH simulations for these setups both with and without the effects of GR.  We integrate all simulations for a total time of $t=3000\,T_{\rm b}$. However, we evaluate the final state of the systems earlier if the total disk mass reaches 40\% of the initial disk mass.  A disk mass lower than this limit results in poor resolution.  While these integration times are less than $t_{\rm GR}$, the relativistic effects on the disk may still be identified.

\begin{figure*}[!ht]
	\includegraphics[width=2\columnwidth]{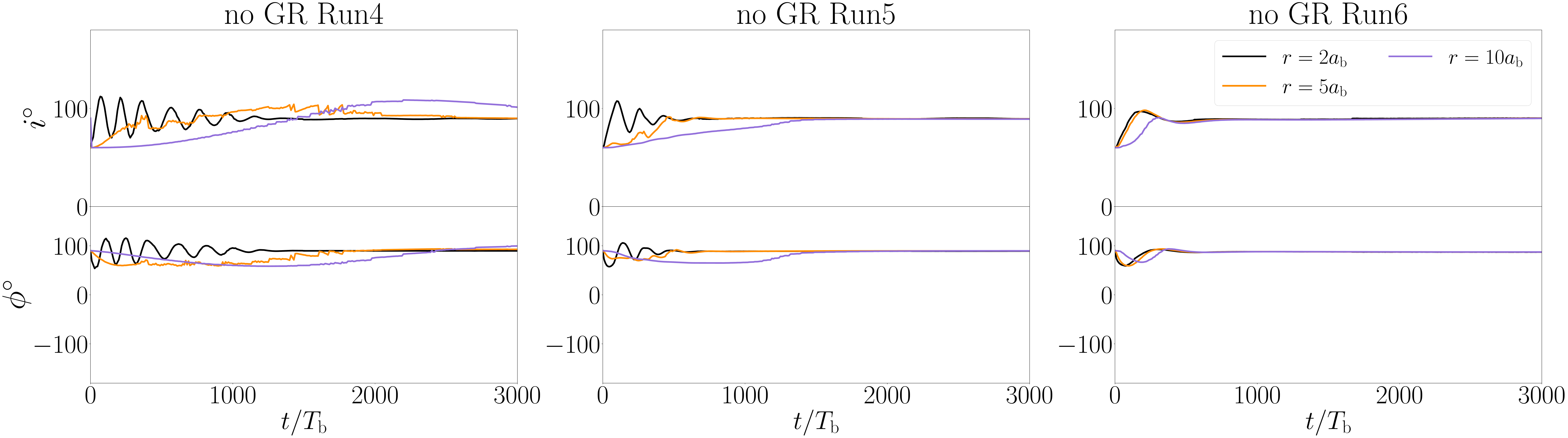}
	\includegraphics[width=2\columnwidth]{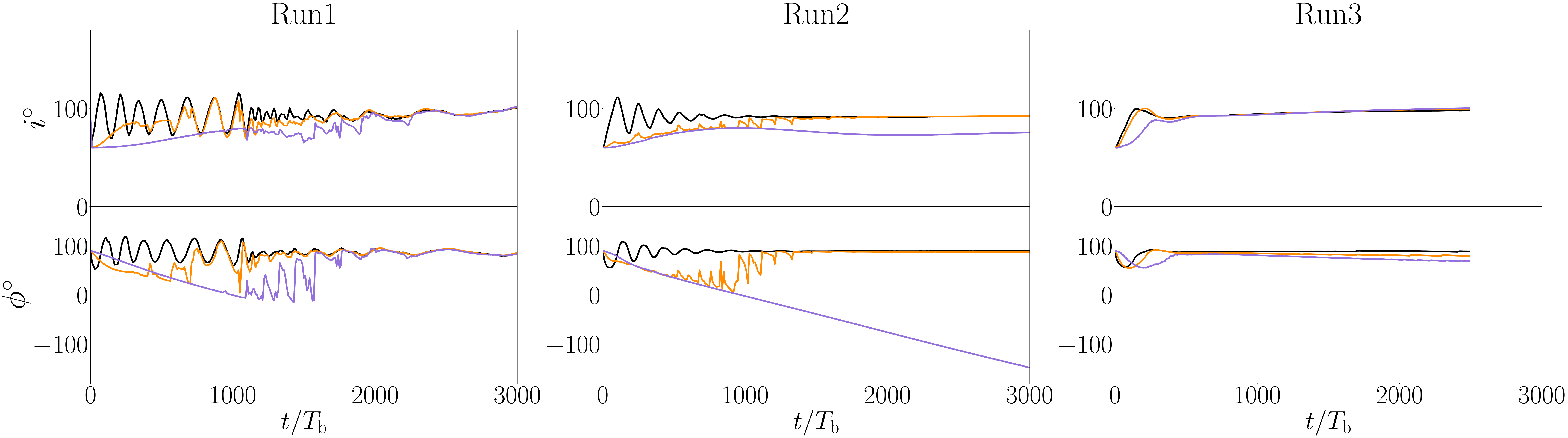}
	\includegraphics[width=2\columnwidth]{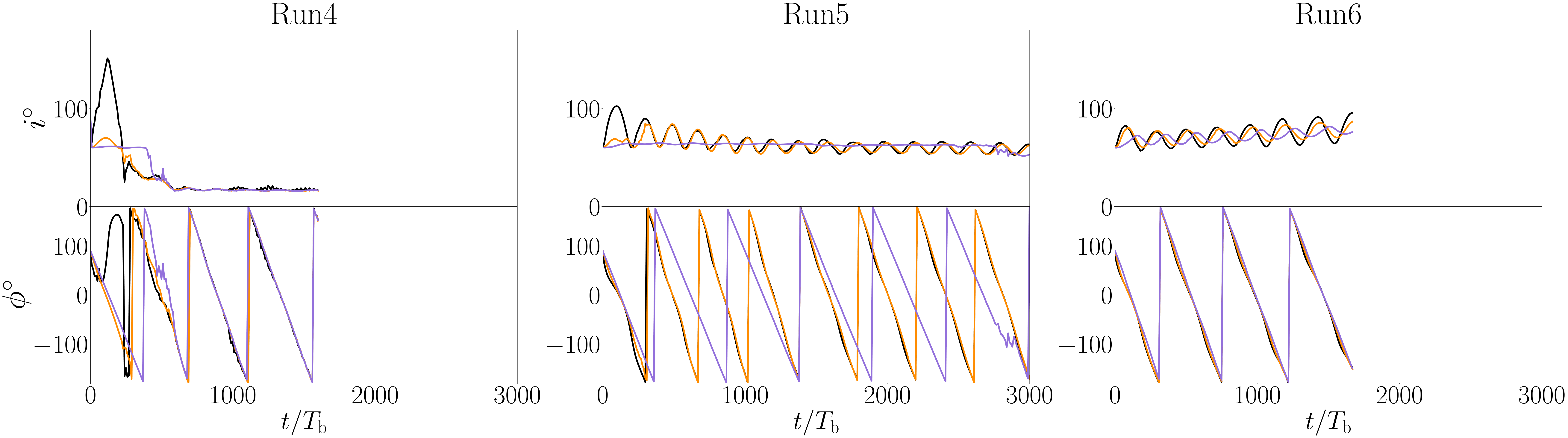}

    \caption{The temporal evolution for the disk inclination ($i$) and nodal phase angle ($\phi$) in the frame of the binary for three different radii in the disk.  The disk inclination is relative to the binary angular momentum vector and the nodal phase angle is relative to the binary eccentricity vector.  In the runs without GR, the binary eccentricity vector is effectively fixed while it rotates about the binary angular momentum vector in the runs with GR.}  In black we show the evolution at $2 \, a_{\rm b}$, orange is at $5 \, a_{\rm b}$, and purple is at $10 \, a_{\rm b}$. The simulations end if the disk mass reaches $40\%$ of the original disk mass.
    \label{fig:timedat}
\end{figure*}

\section{Results}\label{sec:results}

We first describe the simulation results without the effects of GR in which the disk moves to a polar alignment independently of the disk aspect ratio or binary mass.  Then, we examine how GR affects these simulations for two sets of binary parameters as described in Table~\ref{tab:runs}.  While we present results here for a very low mass disk, we expect the behaviour we observe to be unchanged for all circumbinary disks with a mass much less than that of the central binary. 

\begin{figure*}[!ht] \label{fig:gr_splash}
	\includegraphics[width=2\columnwidth]{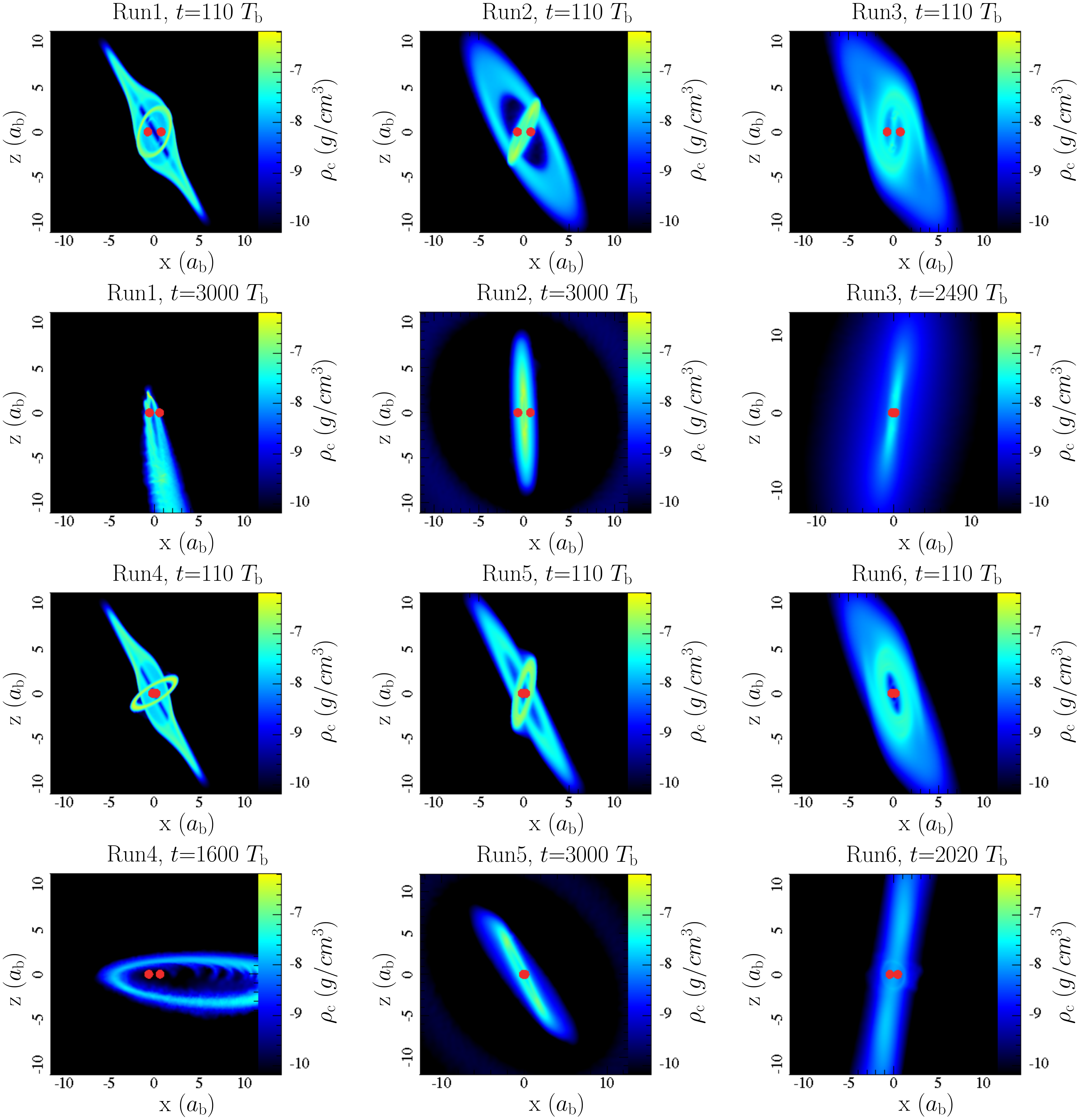}
    \caption{Column density images shown in the $x-z$ plane for the runs with GR at two different times. The binary orbits in the $x-y$ plane and is shown by two red dots.  The binary eccentricity vector is along the $x$-axis.}
\end{figure*}

\subsection{Disk evolution without GR}\label{sec:no_GR}

We first consider the three systems without any GR effects (Run 4, 5 and 6). Note that since the disk mass is very low compared to the binary mass, we expect the same evolution for Runs 1, 2 and 3 as for Runs 4, 5 and 6, in the absence of GR. Since the initial inclination of the disk is large, these disks are all expected to align to polar \cite[e.g.][]{Aly2015, Martin2017, Martin2018, Lubow2018, Zanazzi2018, Cuello2019, Rabago2023}.  

Fig.~\ref{fig:nogr_splash} shows each disk at two different times.  We define a Cartesian coordinate system $(x, y, z)$ in which the binary orbit lies in the $x-y$ plane and the binary eccentricity vector instantaneously lies along the $x$-axis.  The top row shows the system after about $100\,T_{\rm b}$ and  the bottom row shows the systems at $3000 \, T_{\rm b}$. At this time all of the disks have aligned to the stable polar state where they are expected to remain until the gas dissipates.  In all systems we see that the disk initially undergoes a degree of disk warping where the disk plane changes with radius. In all cases we see that the disk becomes sufficiently warped that it breaks into discrete parts \citep{Nixon:2012,Nixon:2013,Dogan:2018,Raj:2021,Drewes:2021}. Finally, in all cases, the whole disk eventually aligns to a polar position.  There is still a small amount of warping in the outer parts of the disk for the lowest disk aspect ratio simulation (Run4).

The top row of Fig.~\ref{fig:timedat} shows the temporal evolution for the disk inclination relative to the binary orbit ($i$) and the nodal phase angle relative to the binary eccentricity vector ($\phi$) for three different radii in the disk.     We see that  all three  disks first undergo nodal libration and tilt oscillations although on different timescales.

The nodal precession timescale for a particle at orbital radius of $3\,a_{\rm b}$ is $157\,T_{\rm b}$ \citep[see equation 16 in ][]{Lubow2018}. This is independent of the black hole mass, it only depends upon the binary mass ratio that we keep fixed. The precession rate of a circumbinary test particle decreases with separation \citep[e.g.][]{Bate2000,Farago2010}.  As a result, the inner disk evolves more quickly but it can be influenced by the outer disk if it is in good radial communication. 
The extent of the radial communication of the inner and outer disks depends on the disk properties \citep[e.g.][]{Nixon2016}.  As a result, there are some differences in the evolution on the way to polar alignment for different disk aspect ratio values.  Larger $H/r$ values result in more effective accretion onto the central binary and more effective radial communication throughout the disk.  As a result, all regions in the disk evolve more quickly to polar in Run6, which has the largest $H/r$ value we simulated.  On the other hand, Run4 has the smallest $H/r$ value and consequently, this disk takes the longest for all regions to evolve polar.  The inner regions of the disks with small $H/r$ undergo tilt oscillations before settling in the polar alignment.

\subsection{Disk evolution with GR}\label{sec:GR}

Next, we consider the evolution of a circumbinary disk around various binary systems with GR.  GR effects lead to prograde apsidal precession of the binary orbit, which, in the frame of the binary, leads to retrograde nodal precession of the disk. Thus, the stationary inclination of the disk increases away from $90^{\circ}$at larger radii.  The critical radius defined in equation~(\ref{rc}) is where the stationary inclination for a test particle reaches $180^\circ$.  
For a critical radius that is much larger than the initial disk outer radius, $r_{\rm c}\gg r_{\rm out}$, the GR precession timescale is long and the disk behaviour is similar to the non-GR case. On the other hand, when the critical radius is small, $r_{\rm c}\ll r_{\rm in}$, then the whole disk precesses on circulating orbits. While breaking may still occur, the eventual disk alignment is coplanar. The timescale for coplanar alignment is much longer than the timescale for polar alignment \citep{Smallwood2019}. We consider two different values for $r_{\rm c}$ in our simulations, both of which are within the radial extent of the disk.


\subsubsection{Larger critical radius}

Runs 1, 2, and 3 have the least massive binary we consider, and thus the largest $r_{\rm c}=8.8\,a_{\rm b}$.  This radius is towards the outer edge of the initial disk surface density. The middle row of  Fig.~\ref{fig:timedat} shows that inner disk librates, the degree to which depending on the disk $H/r$.  Similar to the simulations without GR, this can lead to transient disk breaking. 
The fast radial communication in the simulation with large $H/r=0.1$ (Run3) allows the entire disk to still librate and move to a polar configuration. The outer parts of the disk are farther away than $r_{\rm c}$, but they are connected to the inner polar disk (see  the top two panels of the right hand column in Fig.~\ref{fig:gr_splash}). 
As a result of the more efficient accretion onto the central binary, the disk lifetime is much shorter for systems with larger $H/r$ \citep[e.g.][]{Pringle1981}.

However, for lower disk aspect ratio in Run1 and Run2, the outermost regions of the disk (beyond $r_{\rm c}=8.8 \, a_{\rm b}$ that are shown in purple) decouple from the motion of the binary and undergo slow circulating nodal precession relative to the binary and remain misaligned.  Note that the angles shown in Fig.~\ref{fig:timedat} are relative to the binary orbit and therefore the prograde apsidal precession of the binary orbit driven by GR is seen as a retrograde nodal precession of the disk. In Run2, the nodal precession of the outer disk relative to the binary is on a similar timescale to $t_{\rm GR}$.

For the lowest disk aspect ratio (Run1), the inner librating and outer circulating disks collide. This leads to significant disk accretion onto the binary. The remaining low mass disk is radially narrow, highly eccentric and in a polar alignment as seen in the second image down in the left column of Fig.~\ref{fig:gr_splash}. In the intermediate regime (Run2), the disk remains broken with a polar inner ring and an outer misaligned ring.

\subsubsection{Smaller critical radius}

In Runs 4,5, and 6 the sink particles are more massive and $t_{\rm GR}$ is short enough to drive the disk evolution in a different direction.  The critical radius is $r_{\rm c}=4.6\,a_{\rm b}$, closer to the inner edge of the disk.  In these runs we find that the prograde apsidal precession of the binary is fast enough such that the majority of the disk undergoes circulating nodal precession (see the bottom row of Fig.~\ref{fig:timedat} and the lower two rows of Fig.~\ref{fig:gr_splash}).  In all cases, at least the outer parts of the disk are undergoing retrograde nodal precession on a timescale similar to $t_{\rm GR}=522\,P_{\rm orb}$.

The simulation with the largest disk aspect ratio, Run6, shows a disk that is in good radial communication and undergoing circulating precession on the same timescale at all radii. While the inner disk feels a binary torque that tries to drive nodal libration, the communication from the outer disk is strong. There is little alignment during the simulation because the timescale is much longer than the simulation time.  In linear theory the alignment timescale increases with $H/r$
\citep{King2013}.  For the  large $H/r$ value, efficient accretion leads to a short disk lifetime.  This is the case for Run6 and 40\% of the gas is left after $1670 \, T_{\rm b}$.

Run5 has a lower disk aspect ratio and again, all radii show circulating nodal precession. The disk undergoes breaking early on and remains broken throughout the simulation, but both parts of the disk are circulating. At time $t=3000 \, T_{\rm b}$, 49\% of the initial disk mass remains, the inner disk within $5 \,a_{\rm b}$ has an inclination of $63^{\circ}$, and beyond this region the inclination is $53^{\circ}$.

In the viscous regime in Run4, an inner disk ring initially librates while the outer disk ring circulates. The rings collide leading to significant accretion and only a very low mass disk remains. At $t=1600 \,T_{\rm b}$ the disk is left with only 40\% of its initial disk mass and has an inclination of about $16^{\circ}$ at all disk radii.


\subsection{MS binaries and Stellar-mass black hole binaries}

We have considered simulations with $r_{\rm c}/a_{\rm b}=4.6$ and 8.8. As shown in Fig.~\ref{fig:rc}, these values may be typical for SMBH binaries. However, MS star binaries and stellar-mass black hole binaries likely have larger values. As described in Section~\ref{sec:discussion}, we are unable to simulate the effects of such a large critical radius because of the timescales involved. However, we can predict the outcome based upon our results. Since polar alignment is observed at least in the inner parts of the disk for all of the simulations with $r_{\rm c}/a_{\rm b}=8.8$, we expect that circumbinary disks around binaries with a much larger critical radius will be able to polar align. Therefore polar alignment around MS star binaries and stellar mass black hole binaries is a likely outcome. However, the effects of GR may increase the likelihood of disk breaking if the disk is radially extended enough to be larger than the critical radius. The outer parts of an extended disk in these cases may move toward coplanar alignment.

\section{Conclusions}\label{sec:conclusions}

We modified the SPH code {\sc phantom} to include the prograde apsidal precession of a binary driven by GR and conducted a series of simulations that follow the evolution of a circumbinary gas disk. We considered various disk aspect ratio values in order to examine both the viscous and wave-like disk regimes.   
We also performed a subset of simulations without the effects of GR to better understand how GR drives the evolution of the disk.  In the absence of GR, a circumbinary disk around an eccentric binary can evolve towards  polar alignment for a sufficiently large initial inclination relative to the binary orbit.  However, our results demonstrate that GR can be a destabilizing force in the polar alignment of misaligned circumbinary disks around high mass and/or close separation binaries.   

Whether polar alignment can occur or not depends most strongly upon the value of the critical radius $r_{\rm c}/a_{\rm b}$ given in equation~(\ref{rc}), that describes the radius outside of which there are no librating particle orbits. If $r_{\rm c}$ is smaller than the disk inner radius, then coplanar alignment is the only possible outcome. If $r_{\rm c}$ is larger than the disk outer radius, then polar alignment is possible depending upon the initial disk-binary inclination. For intermediate values, a range of outcomes can take place depending upon the disk properties. For a disk in the wave-like regime, a polar aligned inner disk and a misaligned outer disk is possible. However in the viscous regime, disk breaking can lead to violent collisions and subsequent rapid gas accretion occurs leaving a low mass coplanar disk.

While we are unable to simulate the long timescales required for MS star binaries and stellar-mass black hole binaries, their large $r_{\rm c}/a_{\rm b}$ values  suggest that they are likely to be able to host polar disks, at least close to the binary. Depending on the disk radial extent, at large radii polar alignment may be suppressed meaning that disk breaking is more likely.  Circumbinary disks around SMBHs on the other hand can have much smaller values of $r_{\rm c}/a_{\rm b}$ that can prevent polar alignment.



\begin{acknowledgements}
ACC and AMG acknowledge support from the NSF through grant NSF AST-2107738. RGM and SHL acknowledges support from NASA through grants 80NSSC19K0443 and 80NSSC21K0395. CJN acknowledges support from the Science and Technology Facilities Council (grant number ST/Y000544/1) and the Leverhulme Trust (grant number RPG-2021-380).  SHL thanks the Institute for Advanced Study for visitor support. We acknowledge the use of SPLASH \citep{Price2007} for the rendering of Fig.~\ref{fig:nogr_splash} and Fig.~\ref{fig:gr_splash}.
\end{acknowledgements}

\bibliography{main}{}
\bibliographystyle{aasjournal}

\end{document}